\documentclass[12pt]{article}
\usepackage{epsf}                        
\usepackage{latexsym}                    
\usepackage{amsfonts}                    
\usepackage{amssymb}                     

\newcommand{\sect}[1]{\setcounter{equation}{0}\section{#1}}

\def\be{\begin{equation}}
\def\ee{\end{equation}}
\def\bea{\begin{eqnarray}}
\def\eea{\end{eqnarray}}
\def\nn{\nonumber \\}

\def\hsp#1{\hspace{#1}}

\def\part{\partial}
\def\tfrac#1#2{{\textstyle{\frac{#1}{#2}}}}
\def\half{\tfrac{1}{2}}
\def\x{\times}

\def\incl{\mbox{i}}
\def\Tr{\mbox{Tr}}
\def\STr{\mbox{STr}}

\def\R{\ensuremath{\mathbb{R}}}

\def\makeatletter{\catcode`\@=11}
\makeatletter
\def\mathbox#1{\hbox{$\m@th#1$}}%
\def\math@ccstyles#1#2#3#4#5#6#7{{\leavevmode
      \setbox0\mathbox{#6#7}%
      \setbox2\mathbox{#4#5}%
      \dimen@ #3%
      \baselineskip\z@\lineskiplimit#1\lineskip\z@
      \vbox{\ialign{##\crcr
             \hfil \kern #2\box2 \hfil\crcr
             \noalign{\kern\dimen@}%
             \hfil\box0\hfil\crcr}}}}

\def\mathaccstyles{\math@ccstyles\maxdimen}
\def\maththroughstyles{\math@ccstyles{-\maxdimen}}
\def\unity%
 {\maththroughstyles{.45\ht0}\z@\displaystyle {\mathchar"006C}\displaystyle 1}
\begin{document}

\rightline{KUL-TF/02-15}
\rightline{FFUOV-02/09}
\rightline{hep-th/0212257}
\rightline{December 2002}
\vspace{.4truecm}

\centerline{\Large \bf Non-Abelian Giant Gravitons }
\vspace{.5cm}

\centerline{
  {\bf Bert Janssen${}^{a}$}
 \ \ {\bf and} \ 
  {\bf Yolanda Lozano${}^{b}$}
  }

\vspace{.3cm}
\begin{center}
  {\it ${}^a$ Instituut voor Theoretische Fysica, K.U. Leuven \\
       Celestijnenlaan 200 D, 3001 Leuven, Belgium }\\
  {\tt bert.janssen@fys.kuleuven.ac.be}
\end{center}

\begin{center}
  {\it ${}^b$ Departamento de F{\'\i}sica,  Universidad de Oviedo, \\
    Avda.~Calvo Sotelo 18, 33007 Oviedo, Spain}\\
   {\tt yolanda@string1.ciencias.uniovi.es}
\end{center}
\vspace{.5truecm}

\centerline{\bf ABSTRACT}
\vspace{.5truecm}

\noindent
We argue that the giant graviton configurations known from the literature have 
a complementary, microscopical description in terms of multiple gravitational
waves undergoing a dielectric (or magnetic moment) effect. We present a 
non-Abelian effective action for these gravitational waves with dielectric 
couplings and show that stable  dielectric solutions exist. These solutions
agree in the large $N$ limit with the giant graviton configurations in the 
literature.\footnote{Talk given by Bert Janssen at the RTN workshop in Leuven, 
Belgium in September 2002.}

\sect{Introduction}

It is well-known that a collection of $p$-branes under the influence of a 
background field strength can undergo an ``expansion'' into a single, 
spherical, higher-dimensional $(p+2)$-brane. This is the so-called dielectric 
effect, a first analysis of which was performed in \cite{emparan1}, at the 
level of the Abelian theory relevant to the description of the single expanded 
$(p+2)$-brane. It was some years later that the complementary description from 
the point of view of the lower-dimensional multiple branes was provided
\cite{myers}. From this perspective, the expansion takes place because the
embedding coordinates of the multiple branes are matrix-valued, and give rise 
to new non-Abelian couplings in the combined Born-Infeld-Chern-Simons action 
\cite{TVR, myers}. The macroscopic (Abelian) and the microscopic (non-Abelian)
descriptions have their own range of validity \cite{myers, DTV}, which however 
coincide in the limit where the number $N$ of $p$-branes becomes very large. 
All physical quantities, such as the energy of the configuration and the 
radius of the spherical $(p+2)$-brane take in this limit the same values in 
both descriptions.

The dielectric effect was first derived (under that name) for the case of 
D-branes, but it has become clear since that other types of $p$-branes can 
undergo the same type of effect: the Abelian description given in 
\cite{emparan1} was actually done for fundamental strings and a non-Abelian 
analysis of this effect was provided in \cite{dielF}, in terms of Matrix 
string theory in weakly coupled (linear) background fields. 

The dielectric effect can also occur for multiple coinciding gravitational 
waves. In \cite{DTV} the uplifting of the non-Abelian couplings of D0-branes 
were interpreted as dielectric and magnetic couplings for gravitons in eleven 
dimensions and similar non-Abelian dielectric couplings have been introduced 
in the Matrix model action of the DLCQ description of the pp-wave in eleven 
dimensions \cite{BMN}. A full derivation of the dielectric couplings in the 
effective actions for ten-dimensional gravitational waves, at least up to 
linear order in the background fields, has been given in \cite{JL}. For the 
case of the non-Abelian action for eleven-dimensional waves, the description 
could be easily extended beyond the level of linear approximation. In both 
cases solutions were found of multiple gravitational waves (or gravitons) 
expanding into D2-, D3- and M2-branes.

The question arising then is what is the corresponding, Abelian description of 
this effect. It has been suggested in the literature \cite{DTV, BMN, JL} 
that this might be the configurations of $p$-branes, known as giant 
gravitons. These giant gravitons \cite{ggrav} are spherical M2- and D3-branes 
with non-zero (angular) momentum in Minkowski space or $AdS_m \x S^n$ 
space-times that couple to external flux fields.\footnote{With external flux
field we mean fields that form part of the background in which the brane 
lies and for which the brane is not the source.}
They are stable due to a 
dynamical equilibrium between the tension of the spherical branes that makes 
them contract and the momentum in an external field that makes them expand.

The aim of this letter is to show that indeed the dielectric effect for 
gravitational waves of \cite{JL} and the giant gravitons of \cite{ggrav} 
are two complementary descriptions of the same effect. We will compute
the radius and the energy of the spherical brane configuration and show that 
the values obtained in the microscopic, non-Abelian picture coincide in the 
large $N$ limit with the values in the Abelian, macroscopic picture.

The paper is organised as follows: in section 2 we will review the comparison 
between the microscopical and the macroscopical pictures of the dielectric 
effect for the case of D-branes, since this turns out to be fully analogous to
the case of the microscopic and macroscopic description of giant gravitons. 
In section 3 we give the construction of (Abelian) giant gravitons as given in 
\cite{ggrav}. In section 4 we construct a dielectric M-brane solution from
a non-Abelian action for gravitational waves and compare the large $N$ limit 
of this solution with the Abelian picture and find the the results 
agree up to order $1/N$.

\sect{Macroscopic versus microscopic dielectric branes}

Consider a spherical D2-brane probe laying, say, in the $\theta$ and $\phi$ 
directions of flat ten-dimensional spacetime
\be
ds^2 = -dt^2 + dr^2 + r^2 (d\theta^2 + sin^2 \theta d\phi^2) + dy_{(6)}^2.
\ee
This is, as such, obviously not a stable configuration, since the tension of 
the D2-brane makes the brane contract, while there is no topological 
obstruction to keep the brane at finite radius. However it is possible to 
construct a stable configuration, if we switch on a external flux field and 
make an appropriate choice for the Born-Infeld vector on the D2-brane 
\cite{emparan1, myers}. Taking for the RR 3-form gauge field 
$C_{\mu\nu\rho}$ and the Born-Infeld field strength $F_{ab}$ the following 
values\footnote{Note that this 4-form RR field in flat space is not a 
consistent supergravity background. A proper solution of the dielectric effect
satisfying the supergravity equations of motion, together with an argument
that the probe approximation done here is actually valid, has been given in 
\cite{sugra}.}
\bea
C_{t\theta\phi} &=& -\tfrac{1}{3} f r^3 \sin \theta 
\hsp{.5cm} \Longrightarrow  \hsp{.5cm}
F_{tr\theta\phi} = f r^2 \sin \theta,     \nn
F_{\theta\phi} &=& \half N \sin \theta,
\label{4-form}
\eea 
we notice that the D2-brane is embedded in a constant 4-form flux, while 
the Born-Infeld vector describes the fact that we have $N$ D0-branes 
dissolved in the D2-brane world volume, as can be seen from the Chern-Simons 
action:
\be
S_{CS} \sim T_2 \int_{S^2 \x \R} P[C_1] \wedge F 
       \ = \  N T_0 \int dt \ C_t,
\ee  
where $P[C_1]$ denotes the pullback of the RR-vector $C_\mu$ and we used that 
the tension of the D2 and the D0 are related via
$T_0 = 2 \pi \lambda T_2$.

The full Born-Infeld Chern-Simons action 
\be 
S= -T_2 \int d^3\sigma\  e^{-\phi} \sqrt{\det | g_{ab} + \lambda F_{ab} |}
         \ + \ T_2 \int d^3 \sigma\  P[C_3]
\label{abelian}
\ee
gives then rise to the following potential as a function of the radius $r$ of 
the spherical brane:
\be
V(r) = 2 \lambda^{-1} T_0\Bigl[ \sqrt{r^4 + \tfrac{1}{4}\lambda^2 N^2} 
                   - \tfrac{1}{3} f r^3 \Bigr],
\ee
which for $N \gg 1$ can be expanded as
\be
V(r) = \ N T_0 \ + \ 2 \lambda^{-2} \ T_0  N^{-2} r^4   
       \ - \tfrac{2}{3} \lambda^{-1} T_0 f r^3 \ + \ ... 
\ee
This potential has two extrema, a maximum at $r_1=0$ and a minimum at 
$r_2=\tfrac{1}{4} \lambda N f$, taking the values
\be
V(r_1) = N T_0, \hsp{2cm}
V(r_2) = N T_0 - \tfrac{1}{384}\lambda^2  T_0 N^3 f^4.
\ee 
Clearly, the spherical D2-brane will stabilise at a finite radius $r_2$ and 
has then an even lower energy than if it shrank to zero-size. At this radius,
the contraction due to the D2-brane tension is compensated by the tendency of 
the dissolved D0-branes to expand in the presence of the 4-form flux. The 
spherical D2-brane has no net global charge under the 4-form field strength
(\ref{4-form}), but does have a local charge, which gives rise to a non-zero
dipole moment $\int P[C_3] \neq 0$. Due to this non-zero dipole moment the 
above configuration is called a dielectric brane.

The same effect can also be described from the point of view of the dissolved 
D0-branes. It is well known that a set of $N$ coinciding D-branes exhibits 
an enhanced $U(N)$ symmetry \cite{witten} and should therefore be described by 
a non-Abelian $U(N)$ symmetric effective action and it has been realised that
this non-Abelian action contains a set of non-Abelian couplings to the RR field
of all ranks \cite{TVR, myers}. It is precisely these couplings that give rise
to the dielectric effect. In particular, a set of $N$ coinciding D0-branes in
the presence of an external RR 3-form gauge field is described by the 
Born-Infeld and Chern-Simons actions
\bea
S &=& T_0 \int dt \  \STr \Bigr\{ 
       \sqrt{| P[ E_{tt} + E_{ti} (Q^{-1} - \delta)^{ij} E_{jt}] \det Q |\ {}}
                \  + \   P [ (\incl_X \incl_X) C_3 ]   \  \Bigl \} 
\label{non-abelian}
\eea
where
\be
E_{\mu\nu} = g_{\mu\nu} + B_{\mu\nu},  \hsp{1cm}
Q^i {}_j =\delta^i {}_j + i [X^i, X^k] E_{kj},   \hsp{1cm}
((\incl_X \incl_X) C_3 )_\mu = X^j X^i C_{ij\mu}
\ee
and the $X^i$ are the $U(N)$ matrix valued coordinates transversal to the 
D0-branes. For the static case in a flat background with an external RR 4-form
flux, and after the expansion of the Born-Infeld term and partial integrating 
the Chern-Simons term, this action gives rise to the potential \cite{myers}
\be
V(X) = \lambda^{-2}T_0{\lambda^2}\  \STr \Bigl\{ 
        -\tfrac{1}{4} [X, X]^2 + \tfrac{1}{6}i \lambda [X^k, X^j] X^i  F_{tijk}
           \Bigr\},
\ee
which has a solution to the equations of the form \cite{myers}
\be
F_{tijk} = f \varepsilon_{ijk},  \hsp{1cm}
X^i = -\tfrac{1}{4} \lambda f J^i.
\ee
The $N \x N$ matrices $J^i$ are the generators of the $N$-dimensional 
representation of $SU(2)$, satisfying $[J^i, J^j] = 2i \varepsilon^{ijk} J^k$.
We thus see that the three transverse coordinates $X^i$, whose eigenvalues 
indicate the position of the D0-branes in these directions, span a fuzzy 
(non-commutative) two-sphere with radius
\be
r = \sqrt{\tfrac{1}{N} \Tr ( X^i X^i)} = \tfrac{1}{4} \lambda f \sqrt{N^2 -1}.
\ee 
This can be interpreted as the fact that the coincident D0-branes have, under 
the influence of the RR 4-form, expanded into a fuzzy, spherical D2-brane. 
The energy of this configuration can easily be computed and is given by
\be
E = - \tfrac{1}{384} \lambda^2 T_0 f^4 N (N^2 -1), 
\ee  
which is lower then the zero-energy solution where all $X^i$ commute. Also 
here the 4-form field strength gives rise to a dipole moment, but not to a 
non-zero global D2-brane charge.

Each of these two descriptions, the Abelian macroscopic and the non-Abelian 
microscopic one, have their own range of validity \cite{myers, DTV}. The 
first one is valid when the radius of the spherical D2 is much bigger then 
the string scale, when the higher derivative terms in (\ref{abelian}) are 
neglectable. On the other hand, the non-Abelian calculation is trustable if 
the commutator corrections in higher orders in $F_{ab}$ are small and the 
expansion of the Born-Infeld action (\ref{non-abelian}) converges rapidly, 
i.e.~for the radius much smaller than $\lambda \sqrt{N}$. Clearly both 
conditions are simultaneously met for $N \gg 1$. Indeed we see that the 
expressions for the radii and the energy in both descriptions agree up to 
terms of the order $1/N$.

Another way to understand the agreement of both descriptions is by looking at 
the commutator of the $X^i$:
\be
[X^i, X^j] = \frac{2 i R}{\sqrt{N^2 -1}}\  \varepsilon^{ijk} X^k.
\ee  
In the large $N$ limit the commutators will become very small and the 
non-Abelian character of the microscopic description will diminish rapidly.

\sect{Abelian giant gravitons}

In this section we will review the construction of giant gravitons as 
presented in \cite{ggrav} and draw the attention to the analogies with the 
Abelian description of the dielectric effect.

There are several types of giant gravitons, the most relevant ones being the 
giant graviton in flat spacetime, the genuine giant graviton that lives in the 
spherical part of $AdS_m \x S^n$ and the dual giant graviton, living in the 
$AdS$ part of $AdS_m \x S^n$.  In this letter we will mainly restrict 
ourselves to the giant graviton living in the spherical part of $AdS_7 \x S^4$.
We will comment briefly on the other cases later on.

Consider in eleven dimensions an M2-brane probe wound around a two-cycle 
$\Omega_2$ of radius $L\sin \theta$ in the spherical part of $AdS_7 \x S^4$:
\bea
ds^2 &=& ds^2_{AdS} + L^2 (d\theta^2 + {\cos}^2{\theta}d\phi^2
                        + {\sin}^2{\theta} \ d\Omega_2^2). \nn
d\Omega_{2}^2 &=& d\chi_1^2+{\sin}^2{\chi_1} \ d\chi_2^2.
\eea
Given that there are no non-trivial two-cycles in $S^4$, there is no 
topological obstruction that keeps the M2-brane from contracting to zero 
radius and slipping off the $S^4$. However when the M2-brane carries momentum 
in the transversal direction $\phi$, it will couple to the RR 3-form 
$C_{\phi\chi_1\chi_2}=L^3 {\sin}^3{\theta} \sin{\chi_1}$ 
of $AdS_7 \x S^4$ and will start to expand. A giant graviton is then the 
equilibrium configuration where the contraction by the brane tension $T_2$ is 
canceled by the expansion due to the momentum.  
  
The Hamiltonian of the M2-brane with radius $L\sin{\theta}$ has been computed 
in \cite{ggrav}:
\begin{equation}
\label{potmacr1}
H=\tfrac{P_{\phi}}{L}\sqrt{1+{\tan}^2{\theta}
           ( 1-{\tfrac{{\tilde N}}{P_\phi}\sin{\theta})^2}},
\end{equation}
where ${\tilde N}= 4\pi T_2 L^3$ is an integer that emerges through the 
quantisation condition of the 3-form flux on $S^4$. This Hamiltonian 
(\ref{potmacr1}) has two stable minima, one for $\tan{\theta}=0$, 
corresponding to a point-like object, and another for 
$\sin{\theta}=P_\phi/{\tilde N}$, corresponding to a finite sized 
configuration. The value of the Hamiltonian corresponds in both cases to 
$E=P_\phi/{L}$, i.e.~to a massless particle (hence the name giant graviton). 
Note that the finite radius of the $S^4$ implies an upper bound for the radius
$L \sin \theta$ of the M2 and hence for the momentum: $P_\phi\leq {\tilde N}$.

We can express the total momentum of the giant graviton as $N$ units of 
momentum (gravitons): $P_\phi=N T_0$. In terms of the momentum $T_0$ of the 
gravitons, we find that ${\tilde N}= 2 T_0 L^3 $. 
Therefore we can rewrite the Hamiltonian and its non-trivial solution as
\be
H=\tfrac{NT_0}{L}\sqrt{1+{\tan}^2{\theta} \ 
      ( 1- \tfrac{2L^3}{N}\sin{\theta} )^2 }, \hsp{1.4cm}
\sin{\theta}=\frac{P_\phi}{{\tilde N}}=\frac{N}{2L^3}.
\ee
and the upper bound for the number of gravitons we can put in as 
$N \leq 2 L^3$.

There is a remarkable analogy between the construction of giant gravitons and 
the construction of the spherical D2-brane in the macroscopic picture 
of the dielectric effect. In both cases we are dealing with spherical 
two-branes that are kept stable by a dynamical equilibrium between the 
tensions and the interaction with an external field. For the D2-brane this is 
done via dissolved D0-branes on the world volume, while for the M2 the 
interaction takes place via the angular momentum of the brane. The picture that
arises then is of giant gravitons being M2-branes with dissolved momentum on 
the world volume. In both cases the radius of the D2 and the M2 are 
proportional to the number of dissolved D0's or gravitons.\footnote{In the 
case of the giant gravitons in flat space, the analogy is even more striking 
\cite{DTV, JL}. This is due to the fact that the Abelian D2-brane description 
is in fact the dimensionally reduced version of the flat giant graviton.} 

This analogy raises an obvious question: does there exist also a complementary,
microscopic description of giant gravitons, from the point of view of the 
dissolved gravitons, of which the Abelian description is the large $N$ 
limit? For this to be true, we need a non-Abelian action for multiple 
gravitons (or gravitational waves) with dielectric couplings similar the the 
ones known for D-branes. In the next section we will show that such a 
description does indeed exist.

\sect{Non-Abelian giant gravitons} 

For the case of giant gravitons in flat space, the agreement between the 
microscopic and macroscopic picture has already been pointed out in \cite{DTV},
as an uplifting from the  dielectric D0-brane case in ten dimensions. An 
analysis for the more complicated cases of giant gravitons in $AdS_m \x S^n$
is more difficult, since for this a non-Abelian action for dielectric 
gravitational waves in arbitrary backgrounds is needed.

Such an action was presented in \cite{JL}, making use of Matrix and Matrix 
string theory techniques. In the presence of a 3-form potential $C$, this 
action is given by 
\be
S ={T}_0\int d\tau ~ \STr\Bigl\{ {k}^{-1} \sqrt{
   | P[{E}_{tt}   
   + {E}_{ti}({Q}^{-1}-\delta)^{ij}{E}_{j t}]
\ \det{Q}\ | } \ + \ i  P[(\incl_{X}\incl_{X}){C}]
\Bigr\},
\ee
where
\be
{E}_{{\mu}{\nu}} =
{{\cal G}}_{\mu\nu}
+ {k}^{-1} (\incl_{k} {C})_{{\mu}{\\nu}},
\hsp{1.4cm}
{Q}^i_j =
\delta^i_j + i {k} [X^i,X^k] {E}_{kj}\ .
\label{11non-Abel}
\ee
Note that this action is a gauged sigma model, where the propagation direction
of the gravitational waves appears as a Killing direction which is projected 
out through the effective metric ${\cal G}_{\mu\nu}= g_{\mu\nu} 
- k^{-2} k_\mu k_\nu$ 
and the contraction of the 3-form with the Killing vector 
$(\incl_{k} {C})$ \cite{JL}.

We will not review the construction of this action, but restrict ourselves to
two arguments justifying its validity. First of all, the action 
(\ref{11non-Abel}) reduces to the well-known action (\ref{non-abelian}) for 
coincident D0-branes derived in \cite{TVR, myers}, when reduced along the 
propagation direction of the waves. And secondly, in the Abelian limit, one 
recovers the known effective action for a single graviton. We believe these 
are  non-trivial checks that confirm the interpretation of this action as 
describing non-Abelian M-theory gravitons. 

In order to describe giant gravitons in the spherical part of $AdS_7 \x S^4$, 
we make the following Ansatz for the coordinates on the fuzzy 2-sphere and the 
3-form:
\be
{X}^i=\frac{L\sin{\theta}}{\sqrt{N^2-1}}J^i,
\hsp{2cm}
{C}_{\phi ij}=-\epsilon_{ijk} {X}^k,
\end{equation}
where $J^i,~(i=1,2,3)$ form an $N\times N$ representation of $SU(2)$ and $\phi$
is the propagation direction of the waves. With this Ansatz 
$({X}^i)^2 = L^2{\sin}^2{\theta} \ \unity$. In this particular background 
there is no contribution of the Chern-Simons action. The 3-form potential 
couples however in the Born-Infeld part of the action, through 
$E_{ij}={\cal G}_{ij} + {k}^{-1}{C}_{\phi ij}$, which
implies that the fuzzy 2-sphere carries magnetic moment with respect
to $C$.

Substituting the Ansatz above into the world volume action (\ref{11non-Abel}) 
we obtain the following potential:\footnote{Recall that in our description of 
the gravitons, the propagation direction occurs as an isometry direction, and 
therefore we are dealing with a static configuration, for which we can compute 
the potential as minus the Lagrangian.}
\begin{equation}
\label{potexact}
V(X)= \tfrac{T_0}{L\cos{\theta}}~{\rm STr}\Bigl\{ 
\sqrt{\unity - \tfrac{4L\sin{\theta}}{\sqrt{N^2-1}}X^2 
             + \tfrac{4L^4{\sin}^2{\theta}\cos^2 \theta }{N^2-1} X^2
             + \tfrac{4L^2{\sin}^2{\theta}}{N^2-1}X^2X^2} \ \Bigr\},
\end{equation}
where we have taken into account that some contributions to $\det {Q}$
do in fact vanish after taking the symmetrised average involved in the 
symmetrised trace prescription in (\ref{11non-Abel}). Since we are interested 
in the comparison with the Abelian calculation of section 3, it is convenient 
to look at the large $N$ limit. When $N$ is 
large 
\be
{\rm STr}\{(X^2)^n\} \ \approx \ 
     {\rm Tr}\{(X^2)^n\} + {\cal O} (\tfrac{1}{N^2-1})
  \ = \ L^{2n}{\sin}^{2n}{\theta} N + {\cal O} (\tfrac{1}{N^2-1}),
\label{STr}
\ee
since the commutators involved in the rewriting of (\ref{STr}) can be 
neglected. Therefore, the potential (\ref{potexact}) can be written as:
\begin{equation}
V(\theta) = \tfrac{NT_0}{L\cos{\theta}}
       \sqrt{ 1 - \tfrac{4L^3}{\sqrt{N^2-1}}{\sin}^3{\theta}
                + \tfrac{4L^6}{N^2-1}{\sin}^4{\theta}}
        \  =\  \tfrac{NT_0}{L}
         \sqrt{ 1 + \tan^2{\theta}
                 (1-\tfrac{2L^3}{\sqrt{N^2-1}}\sin{\theta} )^2},
\end{equation}
where we are neglecting terms of order $(N^2-1)^{-\frac{5}{2}}$ in the
expansion of (\ref{potexact}). This potential admits two minima:  the 
point-like graviton at $\sin{\theta}=0$ and the giant graviton at 
\begin{equation}
\sin{\theta}= \half L^{-3} \sqrt{N^2-1},
\label{giantgr}
\end{equation}
both with an energy $E= N T_0 L^{-1} = P_\phi L^{-1}$,  associated to a 
massless particle with angular momentum $P_\phi$. As in the Abelian case, due 
to the finiteness of the radius of the $S^4$ there is again an upper bound on 
the number of gravitons we can add: $\sqrt{N^2 -1} \leq 2 L^3$. 
 
If we compare this non-Abelian computation with the Abelian one of section 3, 
it is clear that the radii, the energy, the upper bound for the angular 
momentum and even the the form of the action agree in the large $N$ limit, up 
to $1/N^2$ corrections. This justifies the statement that the giant gravitons 
of \cite{ggrav} are in fact the macroscopical description of a non-Abelian 
magnetic moment effect for gravitons. 

\vspace{.5cm}
\noindent
{\bf Acknowledgements}\\
The work of B.J. was done as a Postdoctoral Fellow of the Fund for Scientific 
Research-Flanders (F.W.O.-Vlaanderen) and supported in part by the European 
Community's Human Potential Programme under contract HPRN-CT-2000-00131 
Quantum Spacetime.



\begin{thebibliography}{99}

\bibitem{emparan1} R. Emparan, Phys. Lett. B423 (1998) 71, hep-th/9711106.

\bibitem{myers} R. Myers, JHEP 9912 (1999) 022, hep-th/9910053.

\bibitem{TVR}  W. Taylor, M. Van Raamsdonk, 
               Nucl. Phys. B558 (1999) 63, hep-th/9904095. \\
               W. Taylor, M. Van Raamsdonk, 
               Nucl. Phys. B573 (2000) 703, hep-th/9910052.

\bibitem{DTV} S. Das, S. Trivedi, S. Vaidya, JHEP 0010 (2000) 037, 
              hep-th/0008203.

\bibitem{dielF} R. Schiappa, Nucl. Phys. B608 (2001) 3, hep-th/0005145.\\
                P. Silva, JHEP 0202 (2002) 004, hep-th/0111121.\\ 
                D. Brecher, B. Janssen, Y. Lozano, Nucl. Phys. B634 (2002) 23, 
                hep-th/0112180.

\bibitem{BMN} D. Berenstein, J. Maldacena, H. Nastase, JHEP 0204 (2002) 013, 
              hep-th/0202021.

\bibitem{JL} B. Janssen, Y. Lozano, Nucl. Phys. B643 (2002) 399, 
             hep-th/0205254. \\
             B. Janssen, Y. Lozano, {\it A Microscopical Description of 
             Giant Gravitons}, hep-th/0207199.

\bibitem{ggrav} J. McGreevy, L. Susskind, N. Toumbas, JHEP 0006 (2000) 008, 
                hep-th/0003075. \\
                M. Grisaru, R. Myers, \O. Tafjord, JHEP 0008 (2000) 040, 
                hep-th/0008015. \\
                A. Hashimoto, S. Hirano, N. Itzhaki, JHEP 0008 (2000) 051, 
                hep-th/0008016.

\bibitem{sugra} M. Costa, C. Herdeiro, L. Cornalba, Nucl. Phys. B619 (2001) 
                155, hep-th/0105023. \\
                R. Emparan, Nucl. Phys. B610 (2001) 169, hep-th/0105062.  \\
                D. Brecher, P. Saffin, Nucl. Phys. B613 (2001) 218,
                hep-th/0106206.
                
\bibitem{witten} E. Witten, Nucl. Phys. B460 (1996) 335, hep-th/9510135.


\end{thebibliography}
\end{document}